\documentclass[aps,prb,twocolumn,nofootinbib,superscriptaddress]{revtex4-2}
\usepackage{amsfonts, amscd}
\usepackage{amsmath,amssymb}
\usepackage{txfonts}
\usepackage{bm}
\usepackage[dvipdfmx]{graphicx}
\usepackage{physics}
\usepackage{subfigure}

\usepackage{xcolor}
\usepackage[%
  colorlinks=true,
  urlcolor=blue,
  linkcolor=blue,
  citecolor=blue
]{hyperref}

\makeatletter
\let\cat@comma@active\@empty
\makeatother
\begin{document}

\title{
Engineering Yang-Lee anyons via Majorana bound states
}

\author{Takumi Sanno}
\email{sanno@blade.mp.es.osaka-u.ac.jp}
\affiliation{Department of Materials Engineering Science, Osaka University, Toyonaka, Osaka 560-8531, Japan}
\author{Masahiko G. Yamada}
\email{masahiko.yamada@gakushuin.ac.jp}
\affiliation{Department of Physics, Gakushuin University, Mejiro, Toshima-ku, Tokyo 171-8588, Japan}
\affiliation{Department of Materials Engineering Science, Osaka University, Toyonaka, Osaka 560-8531, Japan}
\author{Takeshi Mizushima}
\email{mizushima@mp.es.osaka-u.ac.jp}
\affiliation{Department of Materials Engineering Science, Osaka University, Toyonaka, Osaka 560-8531, Japan}
\author{Satoshi Fujimoto}
\email{fuji@mp.es.osaka-u.ac.jp}
\affiliation{Department of Materials Engineering Science, Osaka University, Toyonaka, Osaka 560-8531, Japan}
\affiliation{Center for Quantum Information and Quantum Biology, Osaka University, Toyonaka, Osaka 560-8531, Japan}
\date{\today}

\begin{abstract}
We propose the platform of a Yang-Lee anyon system which is constructed from Majorana bound states in topological superconductors.
Yang-Lee anyons, described by the non-unitary conformal field theory with the central charge $c=-22/5$, are
non-unitary counterparts of Fibonacci anyons, obeying the same fusion rule, but exhibiting non-unitary non-Abelian braiding statistics.
We consider a topological superconductor junction system coupled with dissipative electron baths, which realizes
a non-Hermitian interacting Majorana system. 
Numerically estimating the central charge, we examine the condition that the non-Hermitian Majorana system can simulate the Ising spin model of the Yang-Lee edge singularity, and  confirm that, by controlling model parameters in a feasible way, 
the Yang-Lee edge criticality is realized.
On the basis of this scenario, we present the scheme for the fusion, measurement and braiding of Yang-Lee anyons in our proposed setup.
\end{abstract} 

\maketitle

\section{introduction}
\label{sec:1}
Anyons are exotic particles which obey unconventional quantum statistics different from fermions and bosons~\cite{WilczekPRL49}, and 
emerge as fractionalized quasiparticles in topological phases of two-dimensional systems 
such as fractional quantum Hall (FQH) states~\cite{TsuiPRL48, LaughlinPRL50, MooreNPB360},
topological superconductors~\cite{ReadPRB61,IvanovPRL86}, and quantum spin liquids\cite{KitaevAP321}.
In last two decades, the application of anyons to fault-tolerant quantum computation, which is referred to as topological quantum computation ~\cite{FreedmanCMP227,freedmanAMS40,KitaevAP321,NayakRMP80,BravyiPRA71,BravyiPRA73,GeorgievPRB74,BravyiIOP12,XuPRA78,XuPRA80,BondersonPRL104,HasslerIOP13,HeckNJP14,HyartPRB88,VijayPRX5,LandauPRL116,AasenPRX6,PluggePRB94,KnappQ2}, has been extensively studied.
In this scheme, informations are stored as topological charges of anyons in a non-local way, and
quantum gates necessary for computation are implemented via the exchange of spatial positions (braiding) of anyons.
In particular, non-Abelian anyons for which the braiding operations are non-commutative unitary operations
acting on a topologically degenerate ground state of many anyon systems are quite useful for the realization of universal topological quantum computation~\cite{KitaevAP321,NayakRMP80}.
One possible candidate of non-Abelian anyons for this purpose is Majorana bound states of topological superconductors~\cite{BravyiPRA71,BravyiPRA73,GeorgievPRB74,BravyiIOP12,IvanovPRL86,BondersonPRL104,HasslerIOP13,HeckNJP14,HyartPRB88,VijayPRX5,LandauPRL116,AasenPRX6,PluggePRB94,KnappQ2,AliceaIOP75,LeijnseIOP27,BeenakkerARCMP4,SatoJPSJ85,MizushimaJPSJ85,MourikScience336,DengNL12,DasNP8,PradaNRP2}.
However, unfortunately, the braiding of Majorana bound states is not sufficient for the construction of universal quantum gates.
More promising candidates are Fibonacci anyons, which enable us to realize universal topological quantum computation only via topologically protected
braiding manipulations~\cite{preskill,BonesteelPRL95,NayakRMP80,KoenigAP325}.
One of the important properties exploited in topological quantum computation is the fusion rule of anyons, which determines what type of anyon appears, when two anyons
are brought together. 
The fusion rule of Fibonacci anyons is,
\begin{eqnarray}
\tau\times\tau=1+\tau,
\label{eq:fusion}
\end{eqnarray}
where $\tau$ denotes the Fibonacci anyon, and $1$ denotes the trivial particle (or vacuum).
This fusion rule determines the Hilbert space where quantum informations are encoded and, combined with braiding, is the basis of universal topological quantum computation.
The fusion rule is also obtained from conformal field theory (CFT) corresponding to the anyon theory.
In general, anyons are described by chiral conformal field theory, which corresponds to gapless chiral edge states at open boundaries of  two-dimensional (2D) gapped topological phases.
In the case of the Fibonacci anyon, the corresponding CFT is the level-1 $G_2$ Wess-Zumino-Witten theory with the central charge $c=14/5$.
There are several proposals for condensed matter systems where Fibonacci anyons appear; e.g. the Read-Rezayi FQH state~\cite{ReadPRB59,CooperPRL87},
superconductor-FQH  junction systems\cite{MongPRX4}, a septuple-layer topological superconductor~\cite{HuPRL120}, and a Rydberg atom gas~\cite{LesanovskyPRA86}.
However, it is still enormously challenging to realize these systems experimentally. 

On the other hand, there is a close cousin of Fibonacci anyon realizable in a non-hermitian quantum system, which is referred to as Yang-Lee anyon~\cite{FreedmanPRB85,ArdonneIOP13}.
Yang-Lee anyons obey the same fusion rule as that of Fibonacci anyons given by Eq.~\eqref{eq:fusion}, and hence, the quantum dimension 
of a Yang-Lee anyon is equal to the golden mean $\phi=(1+\sqrt{5})/2$, which is the measure of the degree of entanglement.
Thus, it is expected that they may have computational power comparable to Fibonacci anyons.
However, Yang-Lee anyons obey {\it non-unitary non-Abelain statistics}; i.e. 
the non-Abelian braidings of Yang-Lee anyon are non-unitary, because of the non-unitary character of the underlying CFT (see Sec.~\ref{sec:2}.~C).
Although this feature is not suitable for the application to unitary quantum computation, it may be utilized for 
the construction of non-unitary quantum gates.
Non-unitary quantum computation has been studied in connection with the measurement-based quantum computation~\cite{TerashimaQIJ3,UsherPRA96,PiroliPRL125,ZhengSci11}.
The systematic realization of non-unitary quantum gates may be useful for simulating non-unitary time-evolution of open quantum systems
in a controllable way. 
The Yang-Lee anyons can  be described by the CFT for Yang-Lee edge singularity with the central charge $c=-22/5$.
The Yang-Lee edge singularity is realized in the one-dimensional (1D) Ising model with an imaginary magnetic field, which is a non-hermitian system. Because of the non-hermiticity,  the corresponding CFT is non-unitary, and the central charge and the conformal weight of a primary field are negative.
It is noted that one can not realize a 2D topological phase in a hermitian system which hosts edge states described by a non-unitary CFT~\cite{FreedmanPRB85}.
However, instead, we can consider a 1D non-hermitian system which effectively simulates the Yang-Lee edge criticality.
Also, 1D anyon models such as the Fibonacci chain model~\cite{FeiguinPRL98}, and the Yang-Lee chain model~\cite{ArdonneIOP13}, 
have the same central charge as the chiral counterparts at criticality, implying the existence of anyons. 

Motivated by these considerations, in this paper, we propose a scheme based on a 1D superconductor junction system which realizes the Yang-Lee edge criticality. Our system consists of topological superconducting nanowire coupled with a semiconductor which plays the role of dissipative electron bath. There are Majorana bound states at open ends of 1D topological superconductors. The coupling with electron baths gives rise to damping of fermion parity for two Majorana bound states, resulting in a non-hermitian term of the effective Hamiltonian. By using the correspondence between Majorana operators and spin operators, we can map this non-hermitian Majorana systems to 
the 1D Ising model with an imaginary magnetic field for the Yang-Lee edge singularity.


As mentioned above, this critical state is not topologically protected, and not suitable for the application to topological quantum computation.
However, our proposal has some advantages. 
In the case of the Read-Rezayi FQH state~\cite{ReadPRB59,CooperPRL87} and superconductor-FQH junction systems~\cite{MongPRX4}, in addition to Fibonacci anyons, 
 there are several different types of quasiparticles which
are described by the $Z_3$ parafermion CFT, and it is quite nontrivial how to detect experimentally the topological charge of Fibonacci anyons, discriminating the Fibonacci anyons from other particles.
In contrast, in the Yang-Lee edge criticality, there is only one type of a nontrivial particle corresponding to a Yang-Lee anyon,
and the field operator for a Yang-Lee anyon is nothing but the scaling limit of the spin magnetization of the non-hermite Ising model,
which is given by the fermion parity of Majorana bound states in the topological superconductor nanowire system.
Thus, the detection and the manipulation of Yang-Lee anyons in the nanowire system is more feasible than Fibonacci anyons in the $Z_3$ parafermion system.

The organization of this paper is as follows. In Sec.~\ref{sec:2}, we summarize important features of the Yang-Lee edge criticality, using the Ising spin model with an imaginary magnetic field, and the CFT analysis. 
We also explain some basic properties of the Yang-Lee anyon model, focusing on the fusion rule and the braiding rules of anyons.
In Sec.~\ref{sec:3}, we present the proposed platform for the Yang-Lee edge criticality based on a topological superconductor nanowire junction system,
which realizes a non-hermitian Majorana many-body system.
In Sec.~\ref{sec:4}, we discuss the stability of the Yang-Lee edge criticality against perturbations which arise from quasiparticle excitations in
the junction system. 
In Sec.~\ref{sec:5}, we present the scheme for realizing and detecting non-unitary non-Abelian statistics of Yang-Lee anyons in our proposed system.
Conclusion is given in Sec.~\ref{sec:6}


\section{Yang-Lee model and Yang-Lee anyon}
\label{sec:2}

\subsection{Yang-Lee model}
In  this section, we briefly review the Yang-Lee Ising spin model which exhibits Yang-Lee edge singularity~\cite{YangPR87,LeePR87}.
The Yang-Lee model $\mathcal{H}_{\rm YL}$ is the transverse 1D Ising model under a pure imaginary magnetic field $ih$:
\begin{align}
\mathcal{H}_{\rm YL} = -\sum^L_{j=1} \left[ J S^z_j S^z_{j+1} + \Gamma S^x_j + ih S^z_j \right].
\end{align}
Here, $S^{x, y, z}_j $ is the spin $1/2$ operator at site $j$, and $L$ is the system size.
Because of the imaginary field, the Hamiltonian is non-hermite.
The Yang-Lee edge singularity is deemed the simplest non-unitary critical phenomenon in non-hermitian systems.
The model exhibits the phase transition characterized by the distribution of zeros of the partition function, 
referred to as the Lee-Yang zeros, on the complex plane of physical parameters~\cite{YangPR87,LeePR87}.
Above the critical field $h_c$, Lee-Yang zeros appear on the line $h>h_c$ in the thermodynamic limit.
In the vicinity of the edge of Lee-Yang zeros, termed Yang-Lee edge singularity at $h_c$, the magnetization exhibits singular behavior.
In 1978, Fisher introduced the non-unitary field theory with the $i \phi^3$ term to explain the Yang-Lee singularity and calculated the critical exponents from a renormalization group method~\cite{FisherPRL40}.
In 1985, Cardy revealed the CFT description of the Yang-Lee edge singularity~\cite{CardyPRL54}. 
The non-unitary CFT for the Yang-Lee edge singularity is the minimal model $\mathcal{M}_{5, 2}$ with the central charge $c=-22/5$ and the scaling dimension $\Delta = -2/5$
for the only one nontrivial primary field.
The CFT description of the Yang-Lee edge singularity in the 2D Ising model with an imaginary magnetic field was confirmed by numerical studies~\cite{ItzyksonIOP2, GehlenIOP24}.
The experimental observation of the Yang-Lee zeros was achieved by measuring quantum coherence of a probe spin coupled to an Ising spin bath~\cite{WeiPRL109,PengPRL114}.

\begin{figure}[htbp]
    \centering
    \includegraphics[width=8cm]{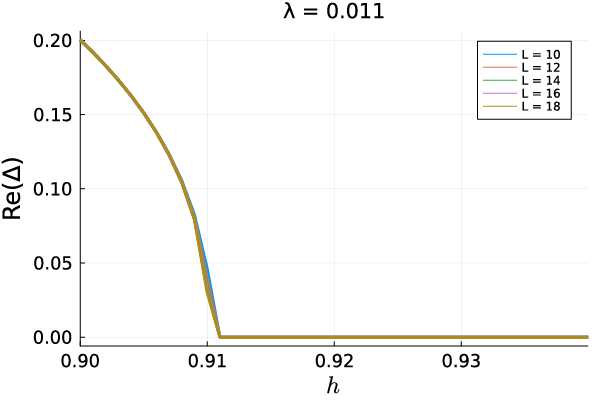}
    \includegraphics[width=8cm]{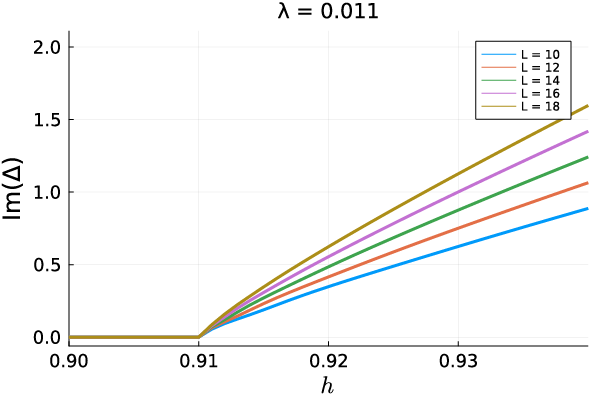}
    \caption{
    The real part and imaginary part of the energy gap $\Delta$ between the ground state and the first excited state for $\lambda = 0.011$ under periodic boundary conditions.
    In the $\mathcal{PT}$-symmetric phase, the real part has finite values and the imaginary part is zero. ${\Delta}$ continuously decreases to zero as
    the system approaches to the critical point.
    In the $\mathcal{PT}$ broken phase, the real part is zero, while the imaginary part has finite values. 
    }
    \label{fig:gap_re}
\end{figure}

\begin{figure}[htbp]
\centering
\includegraphics[width=7cm]{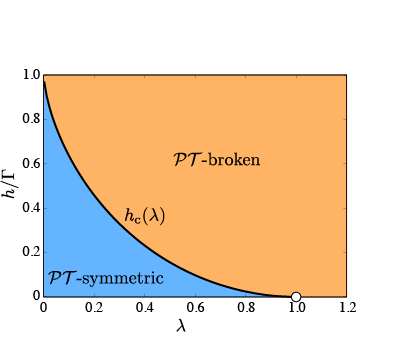}
\caption{The phase diagram of the Yang-Lee model $\mathcal{H}_{\rm YL}$. 
The blue area corresponds to the $\mathcal{PT}$-symmetric phase with real spectra, and the orange area does the $\mathcal{PT}$-broken phase with complex conjugate pair spectra.
The Yang-Lee edge criticality with the central charge $c=-22/5$ is realized on the line of the critical field $h_c(\lambda)$.
}
\label{fig:phase}
\end{figure}

Remarkably, the model possesses $\mathcal{PT}$ symmetry, which imposes an important constraint on the spectrum.
To see this, we introduce the parity operator $P = -i e^{i\pi S^x}$ and the time reversal operator $T= \mathcal{K} \Theta$ where $S^x \equiv \sum S^x_j$, $\mathcal{K}$ is the complex conjugation operator and $\Theta$ is a unitary operator.
In this paper, we select the unitary operator $\Theta$ as the identity operator $\mathbb{I}$.
The parity operator $\mathcal{P}$ flips all spins on the $yz$-plane: 
\begin{eqnarray}
\mathcal{P}: ~(S^x_j, S^y_j, S^z_j) \rightarrow (S^x_j, -S^y_j, -S^z_j).
\end{eqnarray}
The time-reversal operator $\mathcal{T}$ acts on them as,
 \begin{eqnarray}
\mathcal{T}: ~ (S^x_j, S^y_j, S^z_j) \rightarrow (S^x_j, -S^y_j, S^z_j).
 \end{eqnarray}
The Yang-Lee model $\mathcal{H}_{\rm YL}$, therefore, is invariant under $\mathcal{PT}$ transformation and is pseudo-hermitian with respect to the parity operator $\mathcal{P}$.
In general, for pseudo-Hermitian matrices, all eigenvalues are real or complex conjugate pairs, according to the eigenvalue spectrum theory~\cite{BenderPRL80, DoreyIOP34}.
For this reason, the spectrums of the Yang-Lee model is real when $\mathcal{PT}$ symmetry is preserved.
On the other hand, breaking $\mathcal{PT}$ symmetry corresponds to the phase transition from real to complex spectra.
As seen in Fig.~\ref{fig:gap_re}, the drastic change of the ground state energy and the first excitation energy occurs because of $\mathcal{PT}$ symmetry breaking at the critical field $h=h_c$, where we define the parameter $\lambda$ as $\lambda \equiv J/2\Gamma$.
In Fig.\ref{fig:phase}, the phase diagram of the Yang-Lee model $\mathcal{H}_{\rm YL}$ is shown.
In this figure, the blue area is the $\mathcal{PT}$-symmetric phase with real spectra and the orange area is the $\mathcal{PT}$-broken with complex conjugate pairs spectra.
The system shows the Yang-Lee edge criticality with the central charge $c=-22/5$ on the line of the critical field $h_c(\lambda)$.


\begin{figure}[htbp]
    \centering
    \includegraphics[width=8cm]{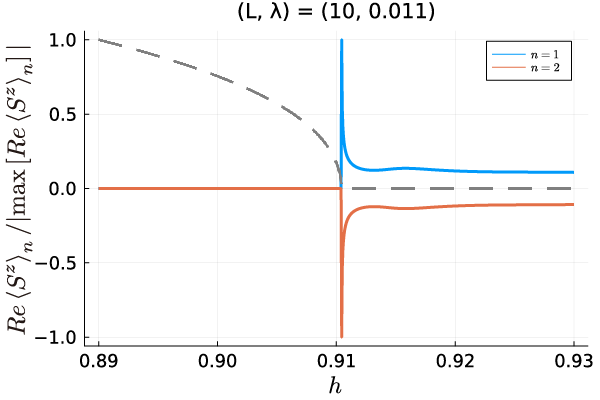}
    \caption{The magnetization $\expval{S^z}_n$ plotted as a function of $h$.
       The blue line $n=1$ (orange line $n=2$) corresponds to the ground state (first excited state) in $\mathcal{PT}$-symmetric phase.
    In the $\mathcal{PT}$-broken phase, the ground state has a negative imaginary part of the spectrum, while the first excited state has a positive imaginary part of that.
    In these numerical calculations, the maximum value of $\abs{\expval{S^z}_n}$ is 13.49 for $n=1, 2$ at $h=0.9105$, which implies the divergent behavior of $\expval{S^z}_n$ at $h=h_c$.
    The dashed line is the real part of the energy gap $\Delta$. 
    }
    \label{fig:sz_bi}
\end{figure}

\begin{figure}[htbp]
    \centering
    \includegraphics[width=8cm]{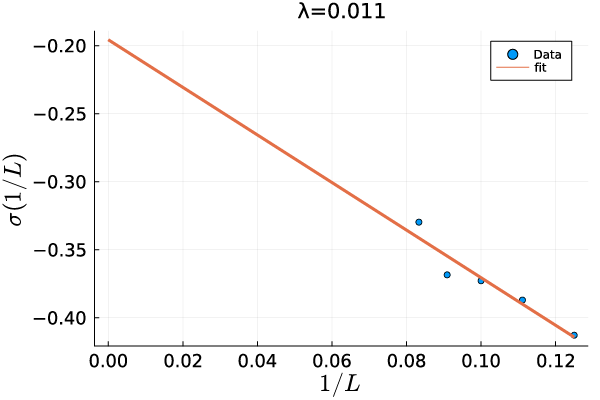}
    \caption{The exponent $\sigma(1/L)$ versus $1/L$.
    The fitting function is the liner function $\alpha + \beta / L$ with
    $\alpha = -0.196$ and $\beta = -1.748$.}
    \label{fig:sz_exp}
\end{figure}

The non-hermitian character of the Yang-Lee model gives rise to unusual critical behaviors.
For instance, in the vicinity of the critical point, the $z$-component of the magnetization exhibits the singular critical behavior with
a negative exponent, 
$\langle S^z \rangle \sim |h-h_c|^{-1/6}$. This divergent behavior is contrasted to conventional critical phenomena where
the order parameter changes continuously from zero to a finite value.
To see this explicitly via numerical calculations, we first note that expectation values of physical quantities in non-hermitian systems are generally expressed with the use of
a biorthogonal basis :
\begin{align}
    &\hat{H} \ket{\varphi_n} = E_n \ket{\varphi_n} \\
   & \bra{\phi_n} \hat{H} = E_n \bra{\phi_n},
\end{align}
where $\ket{\varphi_n} (\bra{\phi_n})$ is a right (left) eigenstate satisfying $\braket{\phi_n}{\varphi_m}=\delta_{n,m}$~\cite{ThomasJMP48,BrodyIOP47}.
The observable value of an operator $\hat{\mathcal{O}}$ can be defined as $\expval{\hat{\mathcal{O}}}_n \equiv \bra{\phi_n} \hat{\mathcal{O}} \ket{\varphi_n}$.
In Fig.~\ref{fig:sz_bi}, we show the magnetization  $\expval{S^z}_n\equiv  \bra{\phi_n} S^z \ket{\varphi_n}$ as a function of $h$ calculated with a biorthogonal basis for the ground state ($n=1$) and the first excited state ($n=2$).
The point where the energy gap $\Delta$ is closed corresponds to the singularity of the magnetization $\expval{S^z}$.
The divergent behavior of the magnetization around $h\sim h_c$ is a signal of the Yang-Lee edge criticality with the negative exponent. 
It is noted that this non-unitary critical behavior can not be obtained by using usual orthonormal systems.
Let us check the exponent of the magnetization by fitting the field-dependence $\expval{S^z}_1 \sim (h-h_c)^\sigma$.
In Fig.~\ref{fig:sz_exp}, we  show the exponent of $\expval{S^z}_1$ as a function $1/L$.
From the fitting data, we found that $ \sigma \approx -0.196 $.
From the renormalization group analysis and the CFT~\cite{FisherPRL40,CardyPRL54}, the exponent $\sigma$ is equal to $-1/6\approx -0.167$ in the thermodynamic limit.
It is noted that, as seen in Fig.~\ref{fig:sz_bi}, the magnetization for the non-trivial ground state and that for the trivial first excited state exhibit
opposite signs. As will be discussed later, this feature can be utilized for the detection of the ground state with the Yang-Lee anyons. 


\subsection{Conformal field theory for Yang-Lee edge singularity}

In this section, we summarize some basics of the CFT for the Yang-Lee edge singularity.
The universality class of the criticality for the Yang-Lee edge singularity is described by the minimal model of CFT.
The minimal model $\mathcal{M}_{p, q}$ contains finite numbers of primary fields, and 
the central charge $c_{p,q}$ and the Kac formula for the conformal weight $h_{r, s}$ are, respectively, given by,
\begin{align}
    &c_{p,q} = 1 - 6 \frac{(p-q)^2}{pq}\\
    &h_{r, s} = \frac{(pr-qs)^2-(p-q)^2}{4pq},
\end{align}
where $p,q$ is the relatively prime number, 
$1 \le r < q, 1 \le s < p$ and its antiholomorphic counterpart $\overline{h}_{r, s}$ takes the same form.
We define the scaling dimension $\Delta_{r, s}$ as $\Delta_{r, s} \equiv h_{r,s} + \overline{h}_{r,s}$.
The Yang-Lee edge criticality is described by the minimal model $\mathcal{M}_{5,2}$
with the central charge $c=-22/5$.
For $p=5,q=2$, there are two primary fields:
\begin{align}
\begin{cases}
\mbox{trivial field} \ \hat{\mathbb{I}} & h_{1,1} = h_{1,4} = 0, \\
\mbox{non-trivial field} \ \hat{\tau} & h_{1,2} = h_{1,3} = -1/5.
\end{cases}
\end{align}
The negative values of the central charge and the conformal weight imply that this is a non-unitary CFT.
For this CFT, there is only one non-trivial primary field, $\hat{\tau}$.
It is noted that $\hat{\tau}$ is the scaling limit of the spin operator $S^z_j$.
The primary field $\hat{\tau}$ satisfies the following fusion rule:
\begin{eqnarray}
\hat{\tau} \times \hat{\tau} \sim \hat{\mathbb{I}}+ \hat{\tau},
\label{eq:fusion2}
\end{eqnarray} 
which is exactly the same as the fusion rule of Fibonacci anyon in Eq.~\eqref{eq:fusion}.
This property is due to the fact that the Yang-Lee anyon theory described by the fusion category is the Galois conjugate of the Fibonacci anyon theory~\cite{FreedmanPRB85,ArdonneIOP13}. 
The fusion rule determines the types of anyons when two anyons
are brought together. 

\subsection{Yang-Lee anyon model}

\begin{figure}[htbp]
    \centering
    \includegraphics[width=9.5cm]{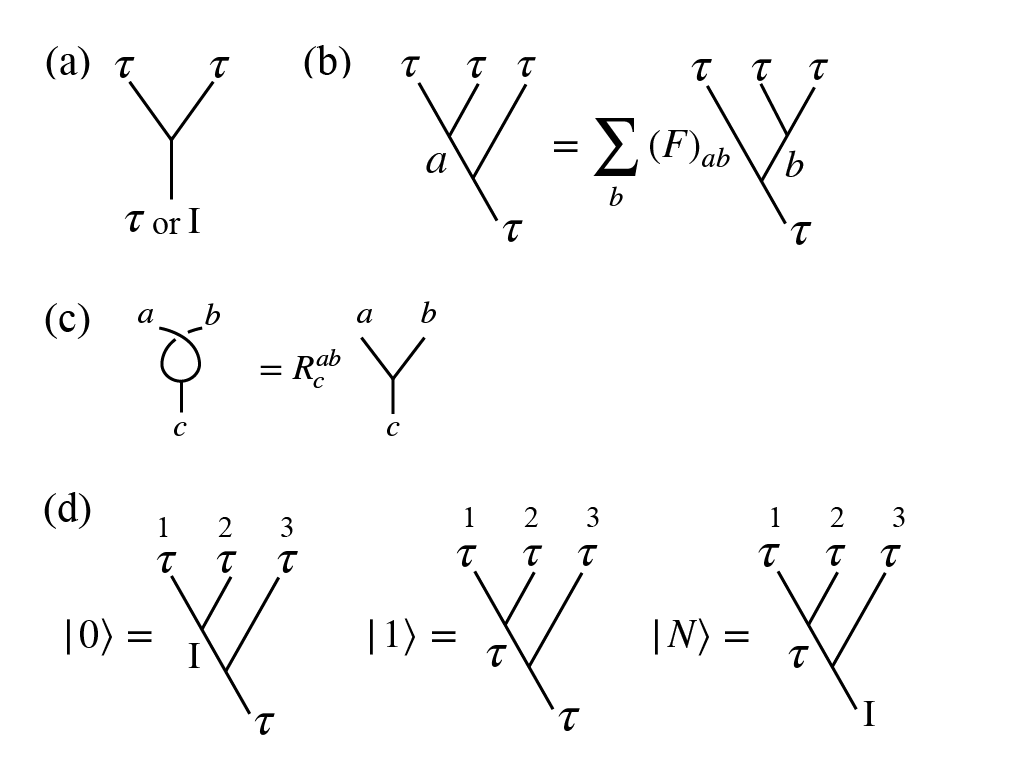}
    \caption{(a) Diagrammatic expression of the fusion rule (\ref{eq:fusion2}). $\tau$ and $\rm I$ denote a state with a Yang-Lee anyon and a vacuum state, respectively.
    (b) Diagrammatic expression of the $F$-symbol. (c)  Diagrammatic expression of the $R$-matrix. (b) Three basis states for non-unitary non-Abelian braiding 
    statistics of Yang-Lee anyons. These states are constructed from three particles labeled as 1, 2, and 3.}
    \label{fig:anyon}
\end{figure}

Here, we briefly explain some basic features of the Yang-Lee anyon model, which are obtained from the modular tensor category~\cite{preskill,KitaevAP321,BonesteelPRL95,HormoziPRB75,NayakRMP80,BondersonPhD}.
Generally, anyons are characterized by anyon statistics different from bosons and fermions.
The braiding rules for anyon statistics are derived from three important factors; the fusion rule, the $F$-symbol, and the $R$-matrix. 
The fusion rule of Yang-Lee anyons is given by Eq.~\eqref{eq:fusion2}, which is also expressed by the diagram shown in Fig.~\ref{fig:anyon}(a).

The $F$-symbol is a matrix for the transformation between different fusion bases, 
which is expressed as shown in Fig.~\ref{fig:anyon}(b). 
The $F$-symbol for Yang-Lee anyons can be derived from that for Fibonacci anyons by changing $\phi \rightarrow -1/\phi$
with $\phi=(1+\sqrt{5})/2$, which corresponds to the Galois conjugation~\cite{FreedmanPRB85,ArdonneIOP13}. 
\begin{eqnarray}
F=\left(
\begin{array}{cc}
-\phi & i\sqrt{\phi} \\
 i\sqrt{\phi} & \phi
\end{array}
\right).
\label{eq:Fs}
\end{eqnarray}
Here, the $F$-symbol acts on the two-dimensional space spanned by the trivial vacuum state and the non-trivial state with a Yang-Lee anyon $\tau$. Note that Eq.~(\ref{eq:Fs}) is a non-unitary matrix, in accordance with the non-unitary CFT for the Yang-Lee edge singularity,
in contrast to Fibonacci anyons for which the $F$-symbol  is unitary.  

The braidings of two anyons are described by the $R$-matrix $R^{ab}_c$, which transposes two anyons $a$ and $b$ fusing into an anyon $c$.
The diagrammatic expression of the $R$-matrix is shown in Fig.~\ref{fig:anyon}(c).
The $R$-matrix is obtained from the topological spin, which describes the phase change arising from the braiding of two anyons, combined with 
the hexagon equation satisfied by the $F$-symbol and the $R$-matrix. 
We, here, conjecture that the topological spin of the Yang-Lee anyons are equal to the conformal spin of the non-trivial chiral primary field $h_{1,2}=-1/5$.
Then, the elements of the $R$-matrix for Yang-Lee anyons are given by,
\begin{eqnarray}
R^{\tau\tau}_{1}=e^{i\frac{2}{5}\pi}, ~~~ R^{\tau\tau}_{\tau}=e^{i\frac{\pi}{5}}.
\label{eq:Rm}
\end{eqnarray}
Eqs.~(\ref{eq:Fs}) and (\ref{eq:Rm}) are the bases for the braiding rule, i.e. the non-Abelian statistics of Yang-Lee anyons.
Following the case of Fibonacci anyons~\cite{preskill,BonesteelPRL95,NayakRMP80}, we consider the three state $|0\rangle$, $|1\rangle$, and $|N\rangle$ depicted in Fig.~\ref{fig:anyon}(d), which are
constructed from three particles labeled as 1, 2, and 3. 
The two states $|0\rangle$ and  $|1\rangle$  are utilized as qubits for the application to quantum computation. 
The state $|N\rangle$ is a non-computation state.
The braiding of the particle $1$ and the particle $2$ is described by the $R$-matrix (\ref{eq:Rm}) alone, and thus,
the matrix for the braiding operation $\sigma_1$ for $1\leftrightarrow 2$  is,
\begin{eqnarray}
\rho(\sigma_1)=\left(
\begin{array}{ccc}
e^{i\frac{2}{5}\pi} & 0 & 0 \\
0 & e^{i\frac{\pi}{5}} & 0 \\
0 & 0 & e^{i\frac{\pi}{5}}
\end{array}
\right),
\label{eq:br1}
\end{eqnarray} 
which acts on the basis states $(|0\rangle,|1\rangle,|N\rangle)^t$.
Next, we consider the braiding operation of the particles $2$ and $3$, denoted as $\sigma_2$.
The matrix for this braiding $\rho(\sigma_2)$ is obtained from the combination of the $R$-matrix and the $F$-symbols as $FRF^{-1}$. 
The result is,
\begin{eqnarray}
\rho(\sigma_2)=\left(
\begin{array}{ccc}
\phi e^{i\frac{3}{5}\pi} & i\sqrt{\phi} e^{-i\frac{\pi}{5}}  & 0 \\
 i\sqrt{\phi} e^{-i\frac{\pi}{5}}  & \phi & 0 \\
0 & 0 & e^{i\frac{\pi}{5}}
\end{array}
\right).
\label{eq:br2}
\end{eqnarray}
Any braiding processes of Yang-Lee anyons are achieved by the combination of $\rho(\sigma_1)$ and $\rho(\sigma_2)$.
It is noted that $\rho(\sigma_2)$ is {\it non-unitary}. Thus, non-Abelian braidings of Yang-Lee anyons give rise to non-unitary transformations on
the qubit states. This property can be utilized for the construction of non-unitary quantum circuits~\cite{TerashimaQIJ3,UsherPRA96,PiroliPRL125,ZhengSci11}, which may be useful
for simulating dissipative quantum dynamics in a controllable way. 
We will discuss how the non-Abelian braiding of Yang-Lee anyons based on Eqs.~(\ref{eq:br1}) and (\ref{eq:br2})
are implemented in our proposed system in the following sections.

\section{Non-Hermitian Majorana System designed for the platform of Yang-Lee edge criticality}
\label{sec:3}

\begin{figure}
\centering
\includegraphics[width=8cm]{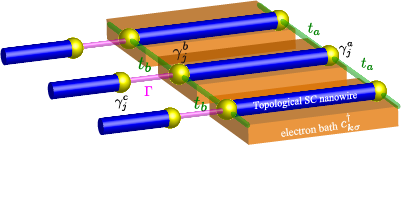}
\caption{Yang-Lee anyon system constructed from Majorana bound states.}
\label{fig:system}
\end{figure}

In this section, we present a scheme of realizing the Yang-Lee edge criticality in heterostructure electron systems.
Our designed system is constructed from topological superconductor (TSC) nanowires coupled with metallic substrates which play the role of electron baths,
as shown in Fig.~\ref{fig:system}.
The key idea here is that the coupling with electron baths gives rise to a non-hermitian term associated with dissipations.
The system consists of coupled units, each of which is composed of two TSC nanowires and a nanoscale metallic substrate.  
In each unit, there are four Majorana bound states at the open edges of the two TSC nanowires.
We assume that one of them is sufficiently away from the other three, and its effect is negligible.
We denote the Majorana fields for the remaining three Majorana bound states as
$\gamma_j^a$,  $\gamma_j^b$, and $\gamma_j^c$.
These three Majorana fields constitute an $s=1/2$ spin operator~\cite{TsvelikPRL69},
\begin{eqnarray}
S^x_j=-\frac{i}{2}\gamma_j^b\gamma_j^c, \quad S^y_j=-\frac{i}{2}\gamma_j^c\gamma_j^a, \quad S^z_j=-\frac{i}{2}\gamma_j^a\gamma_j^b.
\end{eqnarray} 
The Majorana bound states $\gamma^b_j$ and $\gamma^c_j$ are coupled via the tunneling term,
\begin{eqnarray} 
\mathcal{H}_{\Gamma, j}\equiv \frac{1}{2}\Gamma i \gamma^b_j\gamma^c_j=-\Gamma S^x_j,
\label{eq:gamma}
\end{eqnarray}
which simulates the Zeeman interaction due to a transverse magnetic field.
Furthermore, the Majorana bound states $\gamma^a_j$ and $\gamma^b_j$ are coupled to electrons in a nanoscale metallic substrate which plays the role of a dissipative electron bath.
The coupling Hamiltonian is given by,
\begin{align}
\mathcal{H}'_j &= \sum_{k, \sigma} iV^a_{k \sigma j} \left( c^\dag_{k \sigma} + c^{}_{k \sigma} \right) \gamma^a_j + \sum_{k, \sigma} V^b_{k \sigma j} \left( c^\dag_{k \sigma} - c^{}_{k \sigma} \right) \gamma^b_j \nonumber \\
& \qquad +\sum_{k, \sigma} \epsilon_{k \sigma} c^\dag_{k \sigma} c^{}_{k \sigma},
\label{eq:17}
\end{align}
where $c^\dag_{k \sigma} (c^{}_{k \sigma})$ is a creation (an annihilation) operator of electrons with momentum $k$, spin $\sigma$ in the metallic substrate,
$\varepsilon_{k\sigma}$ is its energy band,
 and $V^{a,b}_{k \sigma j}$ are the tunneling amplitudes between Majorana bound states $\gamma_j^a$, $\gamma_j^b$ and the electron bath.
To obtain an effective non-hermitian term corresponding to the imaginary magnetic field of the Yang-Lee model,
we switch to the imaginary-time path integral formulation.
Since Eq.(\ref{eq:17}) is quadratic in electron fields $c^\dag_{k \sigma} (c^{}_{k \sigma})$, we can integrate out these fields exactly.
Then, we arrive at $2\sum_{k, \sigma} V^b_{k \sigma j} \gamma^b_jG_0(\varepsilon_n, k)iV^a_{k \sigma j} \gamma^a_j$ where
$G_0(\varepsilon_n, k)\equiv \frac{1}{i\varepsilon_n-\varepsilon_{k\sigma}}$ is the unperturbed Green function of electrons in the metallic substrate with $\varepsilon_n$ the fermionic Matsubara frequency. We apply the analytic continuation of the Matsubara frequency $i\varepsilon_n \rightarrow \varepsilon +i \delta$ with $\delta$ an infinitesimal quantity, and 
note $G_0(\varepsilon_n, k) \rightarrow {\rm P}\frac{1}{\varepsilon-\varepsilon_{k\sigma}}-i\pi\delta(\varepsilon-\varepsilon_{k\sigma})$.
Assuming that $k$-dependence of $V_{k\sigma j}^{a,b}$ is weak, and that the Fermi energy of the substrate is sufficiently large, and
the density of states of electrons in the substrate is constant, which justify the approximation,
$\sum_{k, \sigma} \rightarrow \int^{\infty}_{-\infty}d\varepsilon_{k\sigma}$, 
we have $\sum_{k, \sigma} V^b_{k \sigma j}V^a_{k \sigma j} {\rm P}\frac{1}{\varepsilon-\varepsilon_{k\sigma}}\approx 0$.
Then, we end up with
 the effective Hamiltonian $\mathcal{H}_{\rm eff}$ with the damping of fermion parity eigenstate:
\begin{align}
\mathcal{H}_{{\rm eff}, j} 
&\sim 2\pi\gamma^a_j \gamma^b_j \sum_{k, \sigma} V^a_{k \sigma j} V^b_{k \sigma j} \delta(\mu - \epsilon_{k \sigma}) \nonumber \\
&=  -ih S_j^z,
\label{eq:imh}
\end{align}
where $h=4\pi\sum_{k, \sigma} V^a_{k \sigma j} V^b_{k \sigma j}\delta(\mu - \epsilon_{k \sigma}) $, and $\mu$ is the Fermi level of the metallic substrate.
Equation~(\ref{eq:imh}) is indeed non-hermitian, and simulates the Zeeman interaction between an imaginary magnetic field and the $z$-component of the spin.
We, furthermore, introduce the tunneling term between Majorana bound states in neighboring units, 
$\gamma_j^{a(b)}$ and $\gamma_{j\pm1}^{a(b)}$:
\begin{eqnarray}
\mathcal{H}_{\rm tun}=\sum_j \left[ i t_a \gamma^a_j \gamma^a_{j+1} + i t_b \gamma^b_j \gamma^b_{j+1} \right].
\label{eq:tunnel1}
\end{eqnarray}
As will be shown below, this term generates the Ising interaction between neighboring spins expressed by Majorana fields.
We deal with $\mathcal{H}_{\rm tun}$ as a perturbation to the eigenstates of the Hamiltonian of the decoupled Majorana units $\mathcal{H}_0=\sum_j[\mathcal{H}_{{\rm eff}, j}+\mathcal{H}_{\Gamma, j}]$.
The eigen energies of each Majorana unit are given by $E_{\pm}=\pm \frac{1}{2}\sqrt{\Gamma^2-h^2}$.
A key assumption in the following analysis is that the hopping amplitudes of $\mathcal{H}_{\rm tun}$ is much smaller than $|E_{\pm}|$,
which justifies the perturbative treatment.
We consider the correction to the Hamiltonian due to the second order perturbation,
$ P\mathcal{H}_{\rm tun}\frac{1}{E_0-\mathcal{H}_0}\mathcal{H}_{\rm tun} P$  where $E_0$ is the eigenenergy of $\mathcal{H}_0$ in the ground state,
$P$ is the projection to the ground state. The intermediate state is the excited state of $\mathcal{H}_0$.
The single-Majorana tunneling process induced by $ \mathcal{H}_{\rm tun}$ results in
 the change of the energy $2E_{+}$.
 The first order correction due to $\mathcal{H}_{\rm tun}$ is suppressed for $t_{a,b} \ll 2E_{+}$.
As a result,  
the second order perturbation with respect to $\mathcal{H}_{\rm tun}$ leads to 
a four-body interaction among Majorana bound states, which simulates the Ising interaction (see Appendix A for more details),
\begin{eqnarray}
\sum_j\frac{J}{4}\gamma_j^a\gamma_j^b\gamma_{j+1}^a\gamma_{j+1}^b=-J\sum_j S^z_jS^z_{j+1},
\label{eq:four}
\end{eqnarray}
where $J\sim t_at_b/E_{+}$. 
Collecting all the terms, we obtain the Majorana Hamiltonian equivalent to the Yang-Lee model,
\begin{eqnarray}
\mathcal{H}_{\rm MF} &=& \sum_j\frac{J}{4}\gamma_j^a\gamma_j^b\gamma_{j+1}^a\gamma_{j+1}^b+\frac{h}{2}\sum_{j}\gamma^a_j\gamma^b_j
+\frac{\Gamma}{2} \sum_j i\gamma^b_j\gamma^c_j  \nonumber \\
& = & \mathcal{H}_{\rm YL}. 
\label{eq:H_eff}
\end{eqnarray}
The above argument implies that $J$ is much smaller than the other energy scales, i.e. $J \ll \sqrt{\Gamma^2-h^2} <\Gamma$.
According to the phase diagram of the Yang-Lee model shown in Fig.(\ref{fig:phase}), in the case of $\lambda=\frac{J}{2\Gamma} \ll 1$, 
the criticality appears for $h\sim \Gamma$.
Thus, the Yang-Lee edge criticality for this parameter region can be realized for the interacting Majorana fermion system (\ref{eq:H_eff}).
For the feasibility of this scenario, it is crucial to suppress the single-Majorana hopping processes in Eq.~(\ref{eq:tunnel1}).
In the following section, we examine the stability of the Yang-Lee edge criticality with the central charge $c=-22/5$ against 
this perturbation (\ref{eq:tunnel1}). 

Before closing this section, we mention about whether the tunnel term (\ref{eq:tunnel1}) affects the $\mathcal{PT}$ symmetry of the Yang-Lee model or not.
We define the $\mathcal{P}$ symmetry and $\mathcal{T}$ symmetry operations of Majorana fields at the $j$-site, which are consistent with those of spin operators as follows:
\begin{eqnarray}
\mathcal{P}&:&  ~~\gamma^a_j \rightarrow   (-1)^j\gamma^a_j, ~  \gamma^b_j \rightarrow   (-1)^{j+1}\gamma^b_j,  
~  \gamma^c_j \rightarrow   (-1)^{j+1}\gamma^c_j,   \nonumber  \\
\mathcal{T}&:& ~~ \gamma^a_j \rightarrow   (-1)^{j+1}\gamma^a_j, ~  \gamma^b_j \rightarrow   (-1)^{j}\gamma^b_j,  ~  \gamma^c_j \rightarrow   (-1)^{j+1}\gamma^c_j,  \nonumber \\
&& i\rightarrow -i.
\label{eq:PTm}
\end{eqnarray} 
Actually, according to these transformation rules,
the Majorana Hamiltonian $\mathcal{H}_{\rm MF}$ preserves the $\mathcal{PT}$ symmetry.
Also, the  tunnel term (\ref{eq:tunnel1}) is indeed invariant under the $\mathcal{PT}$ symmetry operation obtained from Eq.~(\ref{eq:PTm}).
Thus, the perturbations arising from the tunnel term (\ref{eq:tunnel1}) do not break the $\mathcal{PT}$ symmetry of the Yang-Lee model.

\section{Stability of the Yang-Lee edge criticality against perturbations}
\label{sec:4}
Here, we discuss the stability of the Yang-Lee edge criticality against the single-Majorana hopping term (\ref{eq:tunnel1}) via numerical simulations.
In this section, we adopt the periodic boundary condition, because of efficiency of numerical computation.
The universality class is characterized by the central charge $c=-22/5$ and the scaling dimension $\Delta = -2/5$ of the primary field $\hat{\phi}$.
A standard method for the calculation of the central charge is to extract it from the prefactor of the entanglement entropy~\cite{CalabreseIOP06,CalabreseIOP42,BianchiniNPB896,CouvreurPRL119,ChangPRR2,BianchiniIOP48}.
However, a naive application of this approach to
non-hermitian systems occasionally leads to some difficulty in numerics (for details, see Appendix.~\ref{app:2}).
Thus, we exploit the finite-size scaling method ~\cite{BlotePRL56, AffleckPRL56, AlcarazPRL58} combined with the numerical exact diagonalization.

For finite-size scaling, the energy spectra of the Yang-Lee model at criticality are, for the periodic boundary condition, given by,
\begin{align}
& E^{\hat{\mathbb{I}}}(n, \overline{n}) = \epsilon_0 L - \frac{\pi v}{6L}c + \frac{2 \pi v}{L} \left(n + \overline{n} \right)
\label{eq:YLCFT1}
\\
& E^{\hat{\phi}}(n, \overline{n}) = \epsilon_0 L - \frac{\pi v}{6L}c_{\rm eff} + \frac{2 \pi v}{L} \left(n + \overline{n} \right),
\label{eq:YLCFT2}
\end{align}
where we define the effective central charge $c_{\rm eff}$: $c_{\rm eff} \equiv c - 12(h_{1, 2}+\overline{h}_{1, 2})$, and $v$ is 
the velocity of the low-energy excitation.
The indices $n$,  $\bar{n}$ $(\ge 0)$, represent the level of descendant states.  The states with $n=\bar{n}=0$ correspond to the primary states.
For this system, the physical ground state $\ket{\rm GS}$ is the primary state $\ket{h, \overline{h}}$ with the conformal weights $(h, \overline{h})=(-1/5,-1/5)$, and the energy level $E^{\hat{\phi}} (0,0)$.
The first excited state $\ket{\rm 1st}$ is the trivial primary state with the conformal weights $(h, \overline{h})=(0,0)$, which leads to the energy level $E^{\hat{\mathbb{I}}} (0,0)$.
The second excited state $\ket{\rm 2nd}$ is the level $1$ descendant state of the ground state $\ket{\rm GS}$ expressed as,
\begin{align}
\ket{\rm 2nd} \sim L_{-1} \ket{\rm GS} \mbox{~~or~~} \overline{L}_{-1} \ket{\rm GS},
\end{align}
where $L_{-n}$  and  $\overline{L}_{-n}$ ($n=1,2,...$) are the Virasoro generators.
The energy level of the second excited state are given by $E^{\hat{\phi}} (1, 0)$ and $E^{\hat{\phi}} (0, 1)$.
From the energy spectrum obtained by using exact diagonalization, we obtain the central charge $c$, the scaling dimension $\Delta$ and the  velocity $v$.
We set the parameters as $(\lambda, h)=(0.011,0.910),(0.008,0.927)$. (see Appendix \ref{app:1})
We, first, check the Yang-Lee Edge criticality without Majorana hopping term.
From the finite-size scaling and fitting method (Appendix.~\ref{app:1}), we numerically estimate the central charge and the conformal weight (Tab.~\ref{tab:universality_class}).
Numerical results imply that the universality class of the criticality for these parameters coincides with the Yang-Lee edge singularity.
\begin{table}[t]
    \centering
    \begin{tabular}{|c||c|c|} \hline
    $\lambda = 0.011$ & $c$ & $\Delta$ \\ \hline
    {\rm Fit} 1 &$-4.5889$&$-0.4050$\\ \hline
    {\rm Fit} 2 &$-4.5909$&$-0.4072$ \\ \hline
    {\rm Fit} 3 &$-4.6054$&$-0.4106$ \\ \hline
    \end{tabular}
    \centering
    \begin{tabular}{|c||c|c|} \hline
    $\lambda=0.008$ & $c$ & $ \Delta$  \\ \hline
    {\rm Fit} 1 &$-4.4859$&$-0.4278$ \\ \hline
    {\rm Fit} 2 &$-4.5186$&$-0.4272$ \\ \hline
    {\rm Fit} 3 &$-4.5522$&$-0.4264$ \\ \hline
    \end{tabular}
    \caption{Central charge $c$ and scaling dimension $\Delta$. Yang-Lee edge criticality has the central charge $c=-4.4$ and the scaling dimension $\Delta=-0.4$. 
    Fit.1 is a fitting function obtained from data for $L=18,16,14$.
    Fit.2 is that for $12 \le L \le 18$ and Fit.3 is obtained by using all of these data.}
    \label{tab:universality_class}
\end{table}
\begin{figure}
    \centering
    \subfigure{\includegraphics[width=8cm]{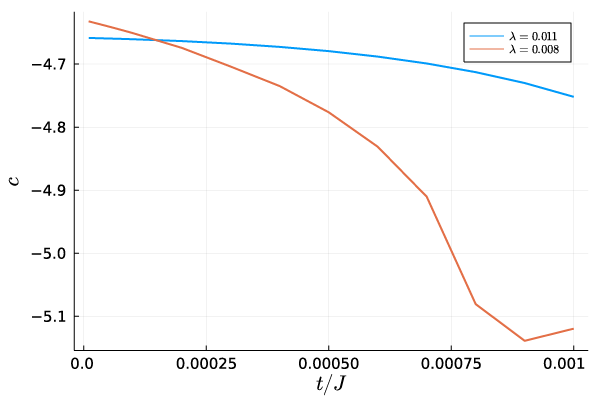}}
    \subfigure{\includegraphics[width=8cm]{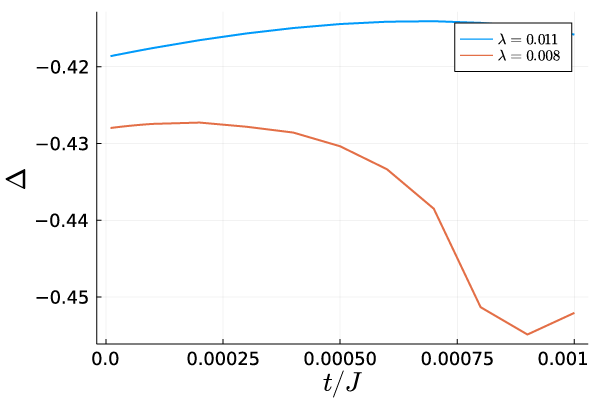}}
    \caption{Central charge $c$ and scaling dimension  $\Delta$ of the nontrivial primary field for the non-hermitian Majorana system with the single Majorana hopping term $\mathcal{H}_{\rm MF}+\mathcal{H}_{\rm tun}$
    plotted as a function of the hopping amplitude $t/J$.
    The blue line is the case of $\lambda=0.011$ and the red line is that of $\lambda=0.008$.}
    \label{fig:class_majorana_hop}
\end{figure}
We, next, examine the stability of the Yang-Lee edge criticality against Majorana hopping term.
In Fig.~\ref{fig:class_majorana_hop}, we show the central charge $c$ and the scaling dimension $\Delta$ versus the amplitude of the 
Majorana hopping term.
It is found that the Yang-Lee edge criticality is stable up to $t/J \sim 10^{-4}$.
The results indicate that the criticality is stable as long as the Majorana hopping term is sufficiently suppressed, which is achieved by
controlling the charging energy $E^{\rm SC}_C$ in our TSC junction system.

\section{Non-unitary non-Abelian statistics of Yang-Lee anyons in the Majorana-based system}
\label{sec:5}

\begin{figure}[htbp]
    \centering
    \includegraphics[width=8cm]{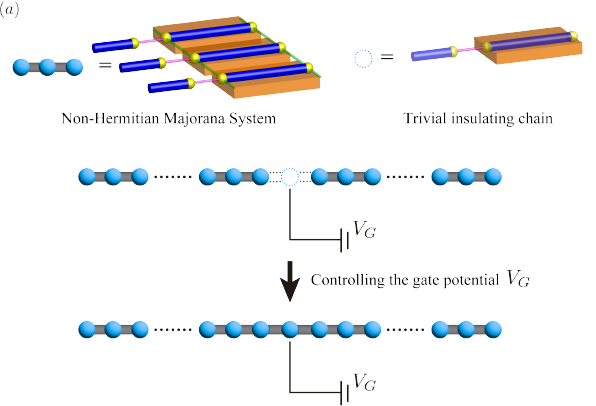}
    \includegraphics[width=8cm]{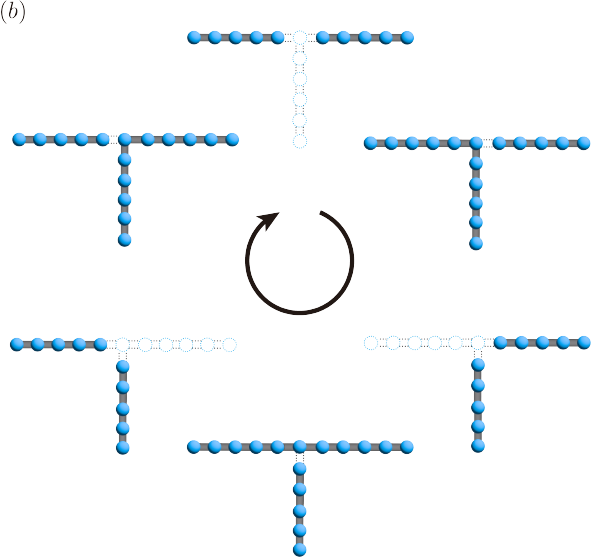}
    \caption{Schematics of a fusion process and a braiding process. 
    (a) Two non-Hermitian Majorana systems are connected with each other through a trivial chain. 
    By controlling the gate potential, this trivial chain can host Majorana bound states.
    Thus, two Yang-Lee chains can be fused into a single Yang-Lee chain.
    (b) By controlling the gate potentials, the region of the Yang-Lee edge criticality can be transferred to other regions, as proposed for the case of Majorana qubits.
    }
    \label{fig:tqc}
\end{figure}

The non-Abelian anyon statistics is detected by the braiding, fusion, and measurement of anyons.
We, here, discuss how to implement these operations of Yang-Lee anyons in our Majorana-based system.
Generally, braidings of anyons are achieved for chiral fields of CFTs.
However, our system is not chiral, but a purely 1D system, which consists of both holomorphic and anti-holomorphic parts of non-trivial fields, if the periodic boundary condition is imposed.
Then, the phase changes of these two parts arising from a braiding operation cancel with each other resulting in trivial braidings only, unless one can manipulate these two parts independently, which is not the case of the Yang-Lee Ising model.
However, on the other hand, in the case of open boundary condition, all the primary fields are expressed  only by the holomorphic part,
and the finite size energy spectrum is given by $E=E_0+\frac{\pi v}{L}(h_{1,2}+ n)$ where $E_0$ is the ground state energy.
Thus, $\tau$ is regarded as a chiral field with a conformal spin equal to $\exp (-i\pi h_{1,2})=\exp (i \pi /5)$.
We conjecture that the nontrivial field $\hat{\tau}$ in this case obeys the rule described by the $F$-symbol (\ref{eq:Fs}) and the $R$-matrix (\ref{eq:Rm}) in Sec.~\ref{sec:2}.
On the basis of this argument,
we consider an array of open chains of the Yang-Lee model which are constructed by using the scheme presented in Sec.~\ref{sec:3}.
In Fig. \ref{fig:tqc}(a), we show the schematic of fusion process.s
The chains are connected with each other through trivial chains which are also constructed in the same scheme, but gate potentials are tuned to
make them in a trivial insulating phase.
By controlling the gate potentials, one can change the trivial chains into a topological state supporting Majorana bound states.
Then, two Yang-Lee chains connected via a trivial chain can be turned into a single Yang-Lee chain, which realizes the fusion of non-trivial Yang-Lee particles.

The detection of the topological charge of a Yang-Lee anyon can be carried out by the measurement of the "magnetization" $S_z$ of the Yang-Lee chain system. As seen in Sec.~\ref{sec:2}. A, the sign of  $\langle S_z \rangle$ is different between the non-trivial ground state with an anyon $\tau$ and the trivial vacuum state (the first excited state). This difference can be utilized for the detection of the state with $\tau$.
Furthermore, the magnetization exhibits a divergent behavior near the critical point as shown in Fig.~\ref{fig:sz_bi} because of the non-unitary feature.
This unique behavior may be  also useful for the experimental detection.
The magnetization $S_z=\frac{i}{2}\gamma_a\gamma_b$ in our Majorana-based system is nothing but a fermion parity, and thus, can be measured by the two-terminal conductance measurement~\cite{PluggePRB94}.

We, finally, comment on how to carry out the braiding operations of Yang-Lee anyons.
We consider two different approaches. 
The first one is the physical transfer of the Yang-Lee anyon state with the use of gate potential control, as proposed for the case of Ising anyons in topological superconductor nanowire systems~\cite{AliceaNP7}.
By changing trivial regions into a topological state, and vice versa, the region of the Yang-Lee edge criticality can be transferred to other regions(Fig.~\ref{fig:tqc}(b)).
Using $Y$-shape junctions, as considered in Ref.~\cite{AliceaNP7}, we can achieve the transposition of two separated Yang-Lee regions
The second approach is to use measurement-based braiding.
The measurement-based braiding was considered before for general anyon systems~\cite{BondersonPRL101}, and also for Majorana systems~\cite{VijayPRB94}. The basic idea is to exploit quantum teleportation induced by projective measurements of topological charges.
Following Ref.~\cite{BondersonPRL101}, we apply a sequence of projective measurements of the topological charge, using the above-mentioned procedure, until we obtain desired results, which leads to the transfer of the state with an anyon.
Applying this measurement-based transfer to two anyons, one can realize braiding of them.

\section{conclusion}
\label{sec:6}

In this study, we design the platform for Yang-Lee anyons which are non-unitary counterparts of Fibonacci anyons, obeying the same fusion rule.
Our proposed systems is a non-hermitian interacting Majorana system constructed from the 1D TSC junctions coupled with dissipative electron baths, which can simulate the Yang-Lee edge criticality.
The coupling with electron baths gives rise to damping of fermion parity in a superconducting nanowire, which play the role of an imaginary magnetic field in terms of the spin representation. 
To stabilize the Yang-Lee edge criticality, it is important to suppress single-Majorana hopping processes in the junction system, which do not exist
in the Yang-Lee spin model.
We can decrease these undesirable processes by controlling the charging energy of TSC nanowires.
We also present the scheme for the braiding, fusion, and detection of Yang-Lee anyons in our TSC junction system.

In the Yang-Lee edge criticality, which is described by the non-unitary CFT with the central charge $c=-22/5$, the ground state
is the non-trivial state corresponding to a Yang-Lee anyon, while the first excited state is a trivial vacuum.
Because of non-unitary character of the CFT, the non-Abelian braiding operator of Yang-Lee anyons is non-unitary, leading to non-unitary non-Abelian statistics.
Although it is not suitable for the application to unitary quantum computation, it may be utilized for simulating non-unitary quantum dynamics in a controllable way.



Our study suggests that non-hermitian Majorana many-body systems embrace rich physics inherent in open quantum systems, and can be 
realized by exploiting topological superconductor junction systems in a feasible and controllable way.
It is a future interesting issue to explore for further exotic phases emerging from non-hermitian Majorana many-body systems.

\section*{acknowledgments}
We would like to thank M. Sato for valuable discussions on non-hermitian quantum systems.
T.S. is supported by a JSPS Fellowship for Young Scientists.
M.G.Y. is supported by Multidisciplinary Research Laboratory System for Future Developments, Osaka University.
This work was supported by JST CREST Grant No.JPMJCR19T5, Japan, and JST PRESTO Grant No.JPMJPR225B, and the Grant-in-Aid for Scientific Research on Innovative Areas ``Quantum Liquid Crystals (JP22H04480)'' from JSPS of Japan, and JSPS KAKENHI (Grant No.JP20K03860, No.JP20H01857, No.JP21H01039, No.JP22K14005 and No.JP22H01221).
A part of the computation in this work has been done using the facilities of the Supercomputer Center, the Institute for Solid State Physics, the University of Tokyo.

\appendix

\section{Derivation of the four-Majorana interaction term}

In this section, we present the details of the derivation of the four-Majorana interaction term Eq.(\ref{eq:four}) in the main text.
The eigen energies of the unperturbed non-hermite Hamiltonian for the $j$-th Majorana unite, $\mathcal{H}_{{\rm eff}, j}+\mathcal{H}_{\Gamma,j}$,
are $E_{\pm}=\pm E$ with $E=\frac{1}{2}\sqrt{\Gamma^2-h^2}$.
The corresponding right eigen states are,
$$
|\phi_{j, R, \pm} \rangle  = u_{\pm} \left |0\right\rangle +v_{\pm} \left|1\right\rangle,
$$
where
 $$u_{+}=\sqrt{\frac{1}{2}\left(1-\frac{ih}{2E}\right)}, \quad v_{+}=-\sqrt{\frac{1}{2}\left(1+\frac{ih}{2E}\right)},$$ 
for the eigen energy $E_{+}$,
 and $u_{-}=-v_{+}$, $v_{-}=u_{+}$ for the eigen energy $E_{-}$,
and $\left |0\right\rangle$ and $\left |1\right\rangle$ are, respectively, the eigen states of the fermion occupation number $\psi_j^{\dagger}\psi_j
=\frac{1}{2}(1+i\gamma^a_j\gamma^b_j)$ for $\psi_j^{\dagger}\psi_j=0 $ and $1$. Here, $\psi_j\equiv (\gamma^a_j+i\gamma^b_j)/2$.
Similarly, the left eigen states are,
$$
\langle \phi_{j, L, \pm} |  = u_{\pm} \left \langle 0\right| +v_{\pm} \left\langle 1\right|.
$$
These states constitute a biorthogonal system, satisfying $\langle \phi_{j, L, s} | \phi_{j, R, s'} \rangle =\delta_{ss'}$ with $s, s'=+, -$, and 
$\sum_{s=+,-} | \phi_{j, R, s} \rangle \langle \phi_{j, L, s} |=1$.
With the use of the biorthogonal basis, we carry out the second-order perturbative expansion in $\mathcal{H_{\rm tun}}$, Eq.(\ref{eq:tunnel1}).
For clarity, we, here, focus on the term $ i t_a \gamma^a_j \gamma^a_{j+1} + i t_b \gamma^b_j \gamma^b_{j+1}$ which acts on the $j$-th and $j+1$-th sites.
we define the right and left eigen states of the unperturbed Hamiltonian for these two sites, respectively, as
$|\phi_{R, s}\rangle =|\phi_{j, R, s_j} \rangle |\phi_{j+1, R, s_{j+1}} \rangle$ and $\langle \phi_{L, s}| =\langle\phi_{j, L, s_j}|  \langle \phi_{j+1, L, s_{j+1}}|$.
Exploiting the relations, $\gamma^a |0\rangle = |1\rangle$, $\gamma^a |1\rangle = |0 \rangle$, $\gamma^b |0\rangle =i |1\rangle$, and 
$\gamma^b |1\rangle =-i |0 \rangle$, we obtain the second-order perturbation correction to the ground state $\ket{\phi_{R, -}}$,
$$
-\frac{t_at_b}{2E}\gamma_j^b\gamma_{j+1}^b\gamma_j^a\gamma_{j+1}^a \ket{\phi_{R, -}}
= -\frac{2t_at_b}{E}S^z_jS^z_{j+1} \ket{\phi_{R, -}},
$$
which leads to the four-Majorana interaction term Eq.(\ref{eq:four}), and hence, the Ising interaction of the Yang-Lee model.
It is noted that the second-order corrections of the order $O(t_a^2)$ and $O(t_b^2)$ give just constant terms.

\section{Method for numerical calculations of the CFT data of the Yang-Lee edge criticality}
\label{app:1}
In this section, we present the details of the method for numerical calculation used in Sec.~\ref{sec:4}.
It is noted that the Yang-Lee edge criticality corresponds to an exceptional point of the non-hermitian Hamiltonian, and thus,
we need to handle numerical diagonalization in the vicinity of the critical point with special care.
We can easily estimate the critical magnetic field $h_c$ from the drastic change of energy spectra at $h_c$.
We check that the critical behavior predicted from the CFT is actually realized at the numerically determined $h_c$ as shown below.

As seen in Fig.~\ref{fig:phase},  the numerical values of $h_c(\lambda)$ are not sensitive to the change of the system size $L$ for $\lambda \sim 0.01$.
On the other hand, $h_c$ strongly depends on the system size for larger values of $\lambda$, e.g., $\sim 1.0$.
For the non-hermitian interacting Majorana model given in Sec.~\ref{sec:4}, there is redundant degeneracy $2^{L/2}$ because of the Majorana representation ~\cite{TsvelikPRL69}, 
and thus, the Hilbert space is much larger than the case of the Yang-Lee spin model, which makes it difficult to carry out numerical calculations for large system sizes.
Because of this reason, we select the parameter $J$ corresponding to $\lambda=0.01$ with $\Gamma=1$, for which the dependence of $h_c$ on $L$
is weak.
We, first, determine $h_c$  numerically from the behavior of the first excitation energy gap $\Delta_1$.
In Fig.~\ref{fig:gap_re}, we show the energy gap $\Delta_1$ for $\lambda = 0.011$.
From this figure, we see that the critical magnetic field $h_c(\lambda)$ is around $h = 0.910$ for this parameter.
For $8 \le L \le 18$, the real and imaginary parts of $\Delta_1$ exhibit drastic changes, which indicates the $\mathcal{PT}$ symmetry breaking around $h=0.910$.

\begin{figure}[htbp]
   \includegraphics[width=8cm]{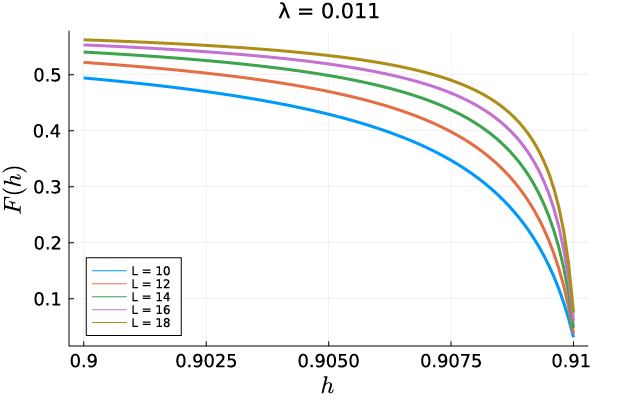}
   \includegraphics[width=8cm]{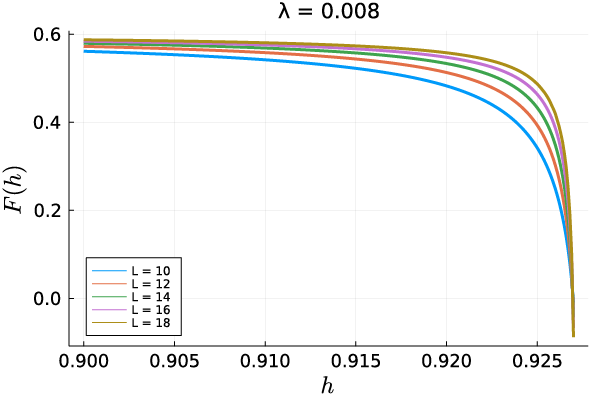}
    \caption{The function $F(h)$ versus $h$}
    \label{fig:check_critical}
\end{figure}

We, next, check for the criticality of the Yang-Lee model as a function of a magnetic field $h$.
We define the function $F(h)$ :
\begin{align}
    F(h) \equiv \frac{E_{1st}(h)-E_{GS}(h)}{E_{2nd}(h)-E_{GS}(h)} + \Delta,
\end{align}
where $E_{GS}$, $E_{1st}$, and $E_{2nd}$ are, respectively, the energy levels for the ground state, the first excited state, and the second excited state, 
and $\Delta=h_{1,2}+\bar{h}_{1,2}$.
The function $F(h)$ must be zero on the Yang-Lee Edge criticality, and is useful for identifying the critical point.
In Fig.\ref{fig:check_critical}, we plot the function $F(h)$ for $0.9 \leq h \leq 0.910$.
From this figure, we can expect that the region in the vicinity of $h=0.910$ is on the Yang-Lee Edge criticality.

We, further, examine the universality class of the criticality by using the finite size scaling method.
The finite size energy spectra of the Yang-Lee CFT are given by Eqs.~\eqref{eq:YLCFT1}-\eqref{eq:YLCFT2} in the main text.
The first step is to fit the spectra with the function form,
\begin{align}
    \frac{E}{L} = a + \frac{b}{L^2},
    \label{eq:fit}
\end{align}
and obtain the velocity $v$ from the relation $2\pi v = b_{\rm 2nd} - b_{\rm GS}$.
The second step is to calculate the central charge $c$ by substituting the velocity $v$ into $- 6 b_{\rm 1st} / \pi v$.
The final step is to estimate the scaling dimension $\Delta$ by using above results for $b_{\rm 2nd} = 2\pi v - \pi v c_{\rm eff}/6$.
\begin{figure}
   \includegraphics[width=8cm]{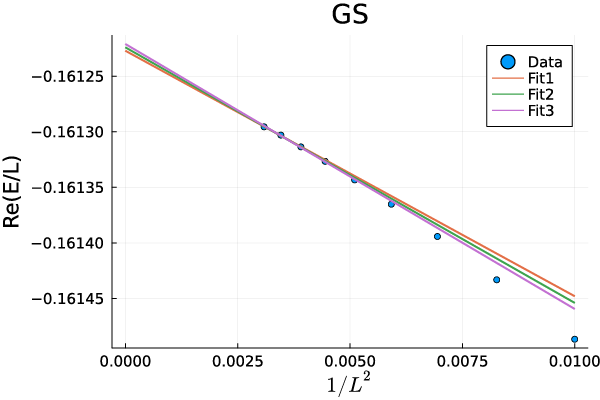}
    \includegraphics[width=8cm]{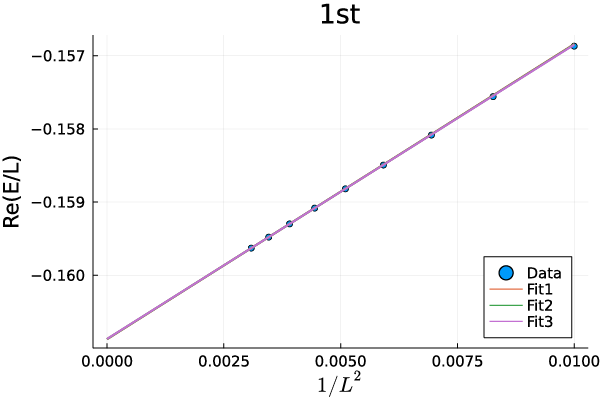}
    \includegraphics[width=8cm]{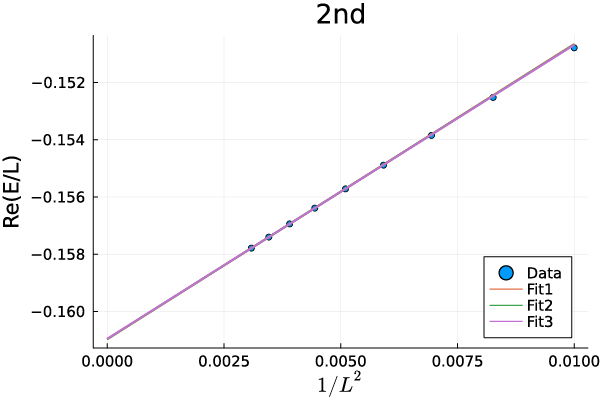}
    \caption{The energy levels of the ground state, the first excited state, and the second excited stat for $\lambda = 0.011$.
    Fit.1 is a fitting function obtained from data for $L=18,16,14$.
    Fit.2 is that obtained from data for $12 \le L \le 18$ and Fit.3 is obtained by using all of them.}
    \label{fig:fit_0.011}
\end{figure}
In Fig.~\ref{fig:fit_0.011}, we show fitting of numerical data of the energy spectra to the function ~\eqref{eq:fit}.
The numerical data well satisfy the finite size energy spectra in Eqs.~\eqref{eq:YLCFT1}-\eqref{eq:YLCFT2}, 
implying that the criticality is described by the Yang-Lee edge CFT.
The results for the CFT data calculated by this method are shown in Table \ref{tab:universality_class} in the main text.


\section{Entanglement Entropy}
\label{app:2}

\begin{figure}[h]
    \centering
    \includegraphics[width=8cm]{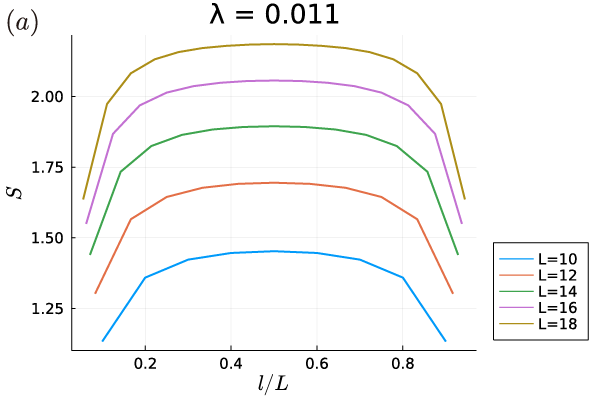}
    \includegraphics[width=8cm]{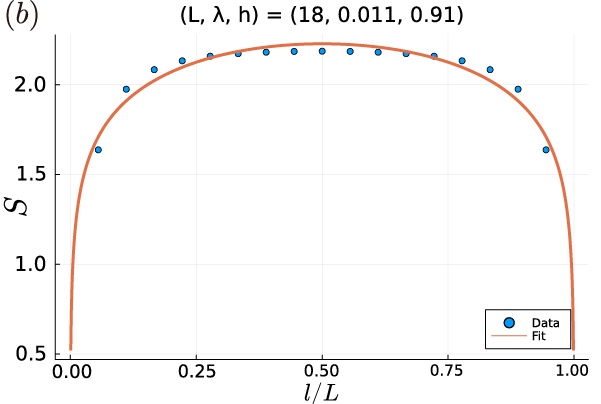}
    \caption{(a) Entanglement Entropy in the vicinity of the Yang-Lee critical point. 
    (b) Entanglement Entropy for $L=18$.
    The fitting function is $\frac{\alpha}{3} \log \left[ \frac{L}{\pi} \sin \left( \frac{\pi l}{L} \right) \right] + \beta$ with $\alpha = 0.889$ and $\beta = 1.711$.}
    \label{fig:EE_1}
\end{figure}

In this section, we examine the Cardy-Calabrese formulae for the entanglement entropy in the case of the Yang-Lee edge criticality.
For the unitary case, the Cardy-Calabrese formulae are quite successful for obtaining the central charge~\cite{CalabreseIOP06,CalabreseIOP42}.
In the case of non-unitary CFTs, it was conjectured that the central charge in the prefactor of the entanglement entropy $S(L)$ for the system size $L$ is replaced with an effective central charge,
\begin{align}
    S(L) = 
        \frac{c_{eff}}{3} \log \frac{L}{\epsilon} + \mathcal{O}(1) 
    \label{eq:EE_nonuni}
\end{align}
where $\epsilon$ is a non-universal ultraviolet cut-off~\cite{BianchiniNPB896,CouvreurPRL119,ChangPRR2}. Here, the periodic boundary condition is imposed.
For the Yang-Lee model, the effective central charge was numerically examined by using Eq.~\eqref{eq:EE_nonuni}~\cite{BianchiniIOP48}.
In the previous study~\cite{BianchiniIOP48}, the entanglement entropy was calculated for the parameter $\lambda=0.9$.
On the other hand, we used much smaller values of the parameter $\lambda\sim 0.01$ to ensure the stability of
numerical calculations against the change of the system size, which is necessary for the numerical calculations in the case of the interacting Majorana model.
For this parameter region, the non-hermitian term, i.e. the imaginary magnetic field, dominates over the Ising interaction term, leading to the situation that
the deviation from the  Cardy-Calabrese formulae for the non-unitary case is serious if the system size is not large enough.
%
To demonstrate this, we calculate the entanglement entropy for the non-hermitian system by using  the biorthogonal basis.
That is, the density matrix $\rho$ is defined as $\rho \equiv \ket{\varphi} \bra{\phi}$, where $\ket{\varphi}$ ($\bra{\phi}$) is the right (left) eigen vector of
the non-hermitian Hamiltonian.
Then, we divid the system into two parts, i.e. system A and system B.
The entanglement entropy of the subsystem A is $S_A = {\rm Tr} \left[ \rho_A \log \rho_A \right]$ where $\rho_A$ is the reduce density matrix of system A by tracing out system B : $\rho_A = {\rm Tr}_B \left[ \rho \right]$.
In Fig.~\ref{fig:EE_1}, we show the calculated results of the real part of the entanglement entropy.
From Fig.~\ref*{fig:EE_1} (b), we estimate the effective central charge $c_{eff}\sim 0.889$, which is inconsistent with the theoretical prediction $c_{eff} = 0.4$.
This slow convergence is due to the logarithmic-dependence on the system size (\ref{eq:EE_nonuni}).
To obtain the correct value of the central charge for this non-hermitian term dominated region, we need to carry out calculations for much larger system sizes,
which requires high numerical costs.
The result indicates that the numerical calculations of the entanglement entropy for non-hermitian systems requires much larger system sizes than
hermitian systems, when non-hermitian terms are predominant.
Because of this reason, we exploit the finite size scaling method for 
the calculation of the central charge in the main text.


\bibliographystyle{apsrev4-1_PRX_style.bst}
\bibliography{yang-lee.bib}
\end{document}